\documentclass{PoS}

\usepackage{graphicx}

\def\epbp{\langle\bar\psi\psi\rangle}

\def\Dslash{\mathop{\not\!\! D}}

\def\LL{\left\langle}   
\def\RR{\right\rangle}  

\def\BE{\begin{displaymath}}
\def\EE{\end{displaymath}}
\def\BEA{\begin{eqnarray*}}
\def\EEA{\end{eqnarray*}}
\def\BNEA{\begin{eqnarray}}
\def\ENEA{\end{eqnarray}}

\PoS{PoS(LAT2005)299}

\title{More evidence of localization in the low-lying Dirac spectrum }

\ShortTitle{More evidence of localization in the low-lying Dirac spectrum }

	\author{C.~Bernard\\ 
	Physics Department, Washington University,
        St. Louis, MO 63130, USA\\
	E-mail: \email{cb@lump.wustl.edu}} 
	\author{Ph.~de~Forcrand\\
        Institute for Theoretical 
        Physics, ETH Z\"urich, CH-8093 Z\"urich, Switzerland\\
	CERN, Theory Division, CH-1211 Geneva 23, Switzerland,\\
	E-mail: \email{forcrand@phys.ethz.ch}}
	\author{Steven~Gottlieb and L.~Levkova\\
        Physics Department, Indiana University,
        Bloomington, IN 47405, USA\\
	E-mail: \email{sg@indiana.edu}, \email{llevkova@indiana.edu}}
	\author{U.~M.~Heller\\
        American Physical Society, One Research Road, Box
        9000, Ridge, NY 11961-9000, USA\\
	E-mail: \email{heller@aps.org}}
	\author{\speaker{J.~E.~Hetrick}\\
        Physics Department, University of the Pacific,
        Stockton, CA 95211, USA\\
	E-mail: \email{jhetrick@pacific.edu}}
	\author{O.~Jahn\\
	Center for Theoretical Physics, MIT,
        Cambridge, MA 02139, USA\\
	E-mail: \email{jahn@mit.edu}}
	\author{F.~Maresca\\
        Physics Department, University of Utah, Salt Lake
        City, UT 84112, USA\\
        E-mail: \email{maresca@physics.utah.edu}}
	\author{D.~B.~Renner and D.~Toussaint\\
        Physics Department, University of Arizona, Tucson,
        AZ 85721, USA\\
	E-mail: \email{dru@physics.arizona.edu},
	\email{doug@physics.arizona.edu}}
	\author{R.~Sugar\\
        Physics Department, University of California, Santa
        Barbara, CA 93106, USA\\
	E-mail: \email{sugar@physics.ucsb.edu}}

\abstract{
We have extended our computation of the inverse participation ratio of
low-lying (asqtad) Dirac eigenvectors in quenched SU(3). The scaling
dimension of the confining manifold is clearer and very near 3. 
We have also computed the 2-point correlator which further
characterizes the localization.
}

\FullConference{XXIIIrd International Symposium on Lattice Field Theory\\
		 25-30 July 2005\\
		 Trinity College, Dublin, Ireland}

\begin{document}

\section{Introduction}

The study of the low-lying eigenmodes of the Dirac operator (LDE),
those corresponding to the smallest eigenvalues, has a rich history since
these modes are thought to be representative of, if not responsible
for, much of the infrared behavior in QCD. Such effects include 
\begin{itemize}

\item Chiral Symmetry Breaking \`a la Banks and
Casher where 
$\epbp \sim \rho_{_\lambda}(0)$

\item The low eigenmodes of $\Dslash$ dominate quark propagators

\item Confinement in many scenarios is thought to be related to 
topological excitations: instantons, monopoles, or
vortices.
These objects all localize Dirac zero-modes in some way.
\end{itemize}

Thus we are interested to learn what we can about this localization, if
indeed it exists, and then to characterize it in some
quantitative way. Hopefully this characterization can tell us
something about the mechanisms responsible for localization.

\section{Inverse Participation Ratio}

The Inverse Participation Ratio (IPR) provides a
quantitative number which characterizes the localization of a scalar
field. For the LDEs, it is defined as
$$
I_i = V \sum_x \rho_i^2(x)
$$
where $V$ is the number of lattice sites $x$, and $i$ labels which
eigenmode is under consideration,
$$
\rho_i(x) = \psi_i^\dagger\psi_i(x).
$$
$\psi_i(x)$ is the $i$-th lowest eigenvector of the (asqtad) Dirac operator
and
$$
\sum_x \rho_i(x) = 1
$$

The IPR takes the following values in cases of different localization:

$\qquad$ $I = 1$ ~~~~if $\rho$ is constant, 

$\qquad$ $I = 1/f$ ~~if $\rho$ is localized (and constant) 
on a fraction $f$ of sites, and 

$\qquad$ $I = V$ ~~~~if $\rho = \delta_{x,x_0}$.

As was first pointed out in \cite{lat04}, the fraction of points
involved in localization should scale with the dimension of the
localizing manifold. For example,
$$
f_{\rm 1-dim} = \frac{{\cal L}/a}{V/a^4}\qquad{\rm or}\qquad f_{\rm 2-dim} =
\frac{{\cal A}/a^2}{V/a^4}, \qquad {\rm etc.}
$$
where in the first, one-dimensional, case ${\cal L}$ is the total length 
of ``localizing material'' and lattice spacing $a$, or 
in the two-dimensional case,
${\cal A}$ is the area of two-dimensional material, and so on for higher
dimensions. The (possibly fractal) 
dimension, $d$, of the localizing manifold is given by the scaling of the
IPR as the lattice spacing is varied,
\begin{equation}
I = 1/f \sim a^{d-4}
\end{equation}

In recent years a growing number of authors have found evidence for
localization of either the LDE or other quantities,
such as the topological charge density \cite{Horvath}. Since the
appearance of \cite{lat04} other groups have analyzed the scaling of
the IPR using improved \cite{Gubarev} and alternative
\cite{Greensite} operators, finding similar conclusions.

While the scaling of the IPR is a rather clean indication of the
underlying localization dimension, its extraction requires a wide
range of lattice spacings to get a fit with good statistics. In
\cite{lat04} for example, the dimension was given as somewhere between
2 and 3. For best results, scaling measurements should be done at {\em
fixed physical volume}, which can then be compared to similar measurements at
{\em fixed lattice spacing} to understand finite size issues.

\section{Scaling of the IPR}

In \cite{lat04} we presented preliminary results for the scaling of
the IPR using quenched lattices (Symanzik 1-loop improved gauge
action) ranging from $12^4$ and $a = 0.2$ fm up to $24^4$ and $a
\sim 0.1$ fm. However, with better statistics on the finest lattices, 
we found that the lattice spacing was
closer to $a = 0.095$ fm, 5\% from of our target of $a = 0.1$ fm.
In order to maintain a fixed volume we
regenerated this ensemble at the target lattice spacing. Furthermore,
as our lattice spacing is set using $r_1$ from the static quark potential,
we use here an updated value of $r_1$ (0.317 fm) taken from
\cite{light} ($r_1 = 0.344$ fm was used in \cite{lat04}). This
increases all lattice spacings by $\sim$10\%, but has no effect on our results.
Additionally, we have increased the convergence criteria for computing 
eigenvectors and added a finer $28^4$ ensemble.

The lattices and parameters used for the present work are collected in
the table below. 
\begin{center}
\begin{tabular}{|c|c|c|c|c|}
\hline
$a$ & $L$ & vol & $\beta$ & no. configs. \\
\hline
0.218 fm~~ & 12 & (2.61)$^4$ fm$^4$ & 7.56  & 100\\
0.163~~~~    & 16 &    (2.61)$^4$   & 7.847 & 100\\
{\em 0.128~~~} & {\em 20} &{\em (2.56)$^4$} & {\em 8.109} & {\em 100}\\
{\tt 0.110~  } & {\tt 24} & {\tt (2.64)$^4$} & {\tt 8.295} & {\tt 100}\\
{\tt 0.0915~  } & {\tt 28} & {\tt (2.56)$^4$} & {\tt 8.527} & {\tt 100}\\
\hline
\end{tabular}
\end{center}
with 64 eigenvectors per lattice. The {\em italicized} ensemble was
rerun for better convergence of eigenvectors (with very little
change), while ensembles in {\tt teletype} are new.

Our main result is summarized in Figure 1, where we plot the scaling
of the average IPR using the lowest 8 eigenvectors.  Compared with
\cite{lat04} we have better statistics on the IPR values and,
particularly, the new value at a smaller lattice spacing (0.0915 fm)
gives a much more precise scaling dimension for the IPR.
\begin{figure}[h]
\begin{center}
\includegraphics[width=400pt]{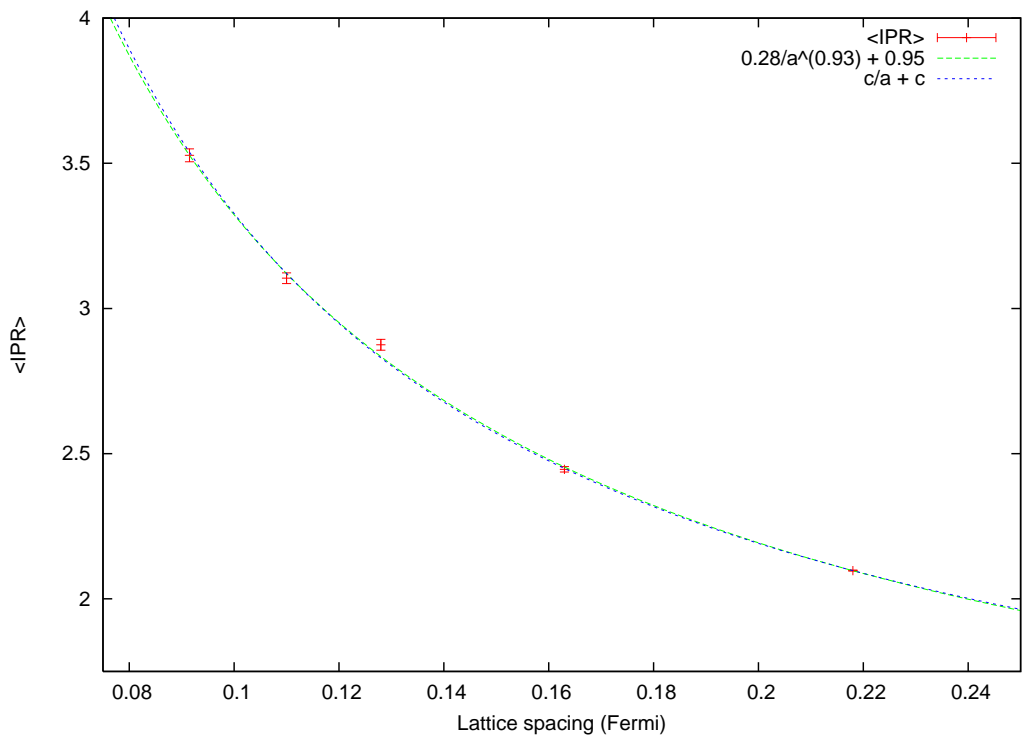}
\end{center}
\caption{The scaling of the IPR for the lowest 8
eigenvectors, in red. In green is the best fit, and associated values.
In blue is a fit to constant + constant/$a$ for comparison.
}
\end{figure}

The best fit to the data has $d-4 = 0.934\pm 0.149$ (see eq. 2.1), thus the
scaling dimension, $d$, of the localization manifold is essentially 3.

\section{Mobility Edge}

In analogy with the mechanism of Anderson localization in condensed
matter physics, we can investigate whether our data shows a {\em
mobility edge}, an  energy above which quarks are delocalized and
below which they show localization. This feature has been investigated
by Golterman and Shamir \cite{GS} and observed in SU(2)
\cite{Gubarev} using an overlap Dirac operator with exact zero
modes. The signal is a reduction in the IPR (less
localization) at some critical value for the eigenvalue--the mobility edge.

In our SU(3) data the IPR values are rather low as compared to IPR
values in SU(2) studies \cite{Gubarev}, \cite{Greensite} (3-4 vs
5-20).
One possible reason for this might be that the localization is due to
topology of one of the SU(2) subgroups, while the other two subgroups
randomize the eigenvectors. Whatever the reason, this is an
interesting clue to understanding the localization.

Only our 28$^4$ lattice at $a = 0.0915$ fm shows a weak indication of the
mobility edge, shown in Figure 2. Its value in physical units, around
50 $\sim$ 100 MeV is consistent with that seen in \cite{Gubarev} . A
study of the IPR on various volumes would be required to confirm this.
\begin{figure}[h]
\begin{center}
\includegraphics[width=400pt, bb=0 50 410 302]{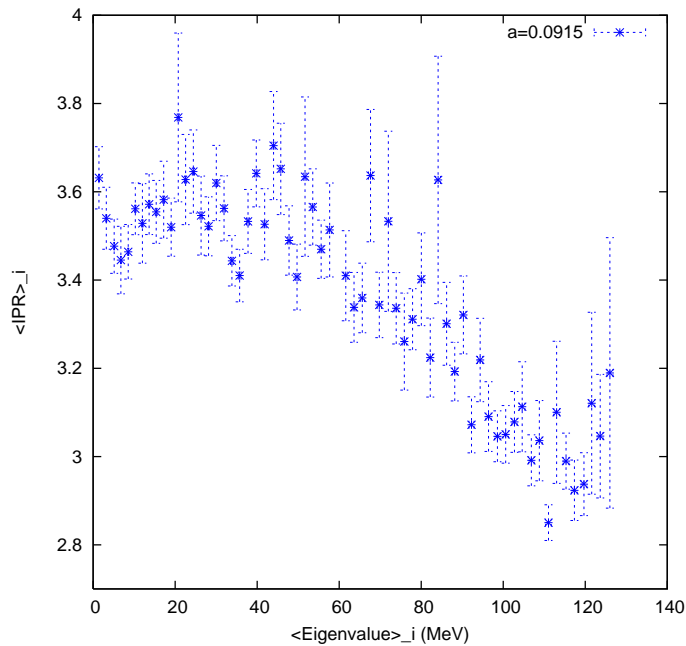}
\caption{$\LL{\rm IPR}\RR$ vs eigenvalue in physical units.
The mobility edge is weakly indicated by
 a decrease in IPR at about 60$-$100 MeV.}
\end{center}
\end{figure}

\section{Two-point Correlations}

While the IPR is a good quantitative indicator of localization, it
only tells us the fraction of lattice sites where the eigenvector is
large. If we rearranged the lattice sites we would obtain the same
IPR. The two-point correlator on the other hand gives information on the
connectedness of the eigenvector, and should drop off as $\sim
r^{d-4}$ if the eigenvector is localized uniformly on a $d$-dimensional
manifold.

We have computed the ``all-to-all'' correlator, $\LL\rho(x)\rho(y)\RR$
and present the average at different spatial separation $|x-y|$ for
eigenvalue 0 in Figure 3. These correlators have parity sawtooth
behavior familiar in staggered fermion propagators, which lessens as
the lattice spacing decreases. Taking the slope of the correlator from
the even points, we show a line with $d=3.5$ (a very similar result if
achieved for the odd points which lie slightly below the line).
\begin{figure}[h]
\begin{center}
\includegraphics[width=400pt]{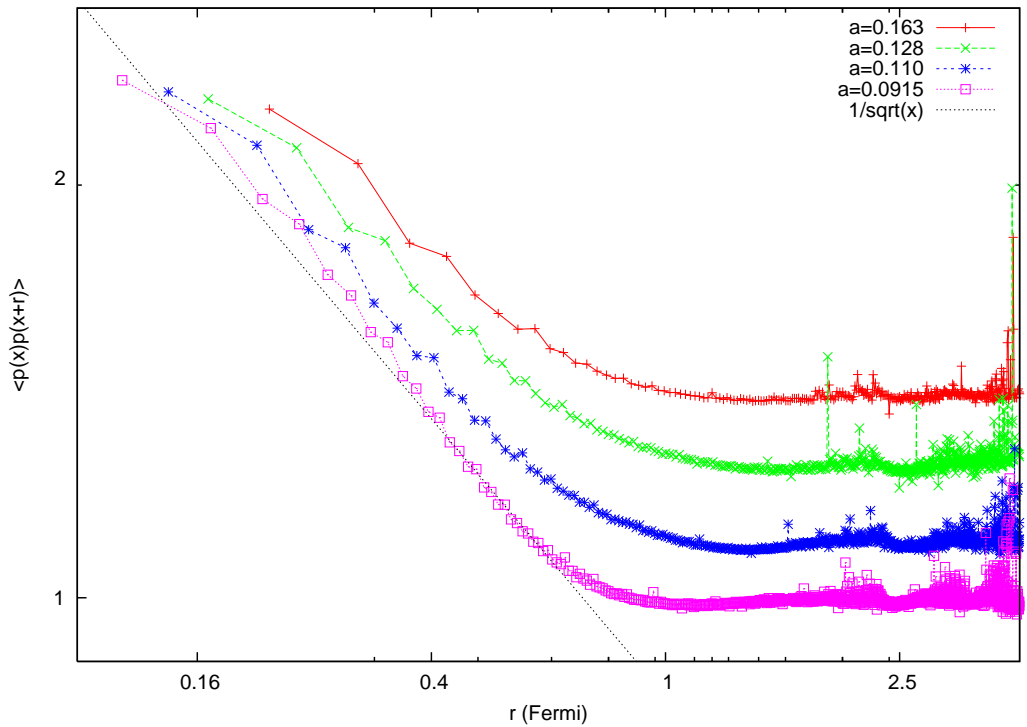}
\caption{The two-point correlator of the lowest eigenvector, 
$\sum_x \LL\rho(x)\rho(x+r)\RR$ as a function of distance $r$.
Plots (except a=0.0915) are displaced upward for clarity. The axes are 
log-log.}
\end{center}
\end{figure}
We see that in physical units, the correlation disappears by about 1
fermi, for all of our lattices.


\section{Conclusions}

We have extended our study of the IPR of the lowest eigenvectors of
the asqtad Dirac operator computed on quenched background gauge
fields. A number of conclusions are apparent.

\begin{itemize}

\item The IPR scaling $\sim 1/a$ implies a dimension $d = 3$ for 
the localizing manifold. The signal is quite a bit clearer than
in our previous study.

\item We see a weak mobility edge in our finest lattices (only)
at 50$\sim$100 MeV. This is consistent with other work. The fact that
our signal is weak is likely attributed to the lack of exact zero
modes of our Dirac operator and possibly to the larger gauge group
(SU(3) vs SU(2)).

\item The two point correlator $\LL\rho(x)\rho(x+r)\RR$ suggests a
fractal dimension of $\sim 3.5$. It should be noted that we have
not finished computing this quantity on all of our largest lattices
(Figure 3 represents only about 8 of our 28$^4$ dataset, however the
correlator shows little fluctuation from lattice to lattice). In
physical units the correlation disappears at $\sim$1 fm.

\end{itemize}

Our findings do not support the naive picture where the low-lying Dirac
eigenmodes are localized on monopoles ($d$=1) or vortices ($d$=2). The
relationship between topological excitations and eigenvector localization
is more subtle. For instance, a set of center vortices gives eigenvectors
localized weakly on the vortices themselves, and more strongly on their
intersections \cite{Reinhardt}.

\end{document}